\documentstyle[sprocl,epsfig]{article}

\newcounter{figureno}
\newcommand{\capttwo}[2]{
\phantom{mmmm}
\vspace*{6mm}
\parindent=0pt
\renewcommand{\baselinestretch}{0.5}
\begin{minipage}[t]{130mm}
\mbox{
\addtocounter{figureno}{1}
\begin{minipage}[t]{60mm}
\small\sl Figure~\thefigureno.\
#1
\end{minipage}}
\phantom{i}
\mbox{
\addtocounter{figureno}{1}
\begin{minipage}[t]{60mm}
\small\sl Figure~\thefigureno.\
#2 
\end{minipage}}
\end{minipage}
\vspace*{-5mm}}
\def\Journal#1#2#3#4{{#1} {\bf #2}, #3 (#4)}

\def\NIMA{{\em Nucl. Instrum. Methods} A}

\def\PLB{{\em Phys. Lett.}  B}
\def\PRL{\em Phys. Rev. Lett.}
\def\PRD{{\em Phys. Rev.} D}
\def\ZPC{{\em Z. Phys.} C}
\def\ra{\rightarrow}

\def\be{\begin{equation}}
\def\ee{\end{equation}}
\def\bea{\begin{eqnarray}}
\def\eea{\end{eqnarray}}
\def\b0{$B^0_d$}
\def\b+{$B^+$}
\def\b0s{$B^0_s$}
\def\b*0{$B^{*0}_d$}
\def\b*+{$B^{*+}$}
\def\b*0s{$B^{*0}_s$}
\def\b*{$B^\star$}

\begin{document}

\title{REVIEW OF b-FLAVORED HADRON SPECTROSCOPY}

\author{ G. Eigen }

\address{Department of Physics, University of Bergen,
Allegaten 55, 5007 Bergen, Norway}

\maketitle\abstracts{
The current status of b-flavored hadron spectroscopy is reviewed.}
  
\section{Introduction}

B-flavored hadrons are the heaviest flavored hadrons, since due to $V_{tb}\sim 1$
the top quark decays long  before forming bound states. They are copiously produced in 
$ p \bar p$ collisions at the Tevatron and in $Z^0$ decays at LEP. 
Each LEP experiment has collected $ 8.8 \times 10^5 \ b \bar b$ events.
The formation of B-mesons versus b-baryons at the $Z^0$ is favored by 9:1  
The heavy quark $(\bar Q)$ also dresses favorably with  light quarks produced
in soft gluon processes. This leads to production rates of
$ B^+_u:B^0_d:B^0_s =1 : 1 : {1 \over 3}$. $B_c$ production is reduced  
by $2 - 3$ orders
of magnitude, since a hard gluon process is needed. 
  
The properties of heavy-light systems ($ \bar Q q$ or $  Q  q  q$) 
are predicted by heavy quark effective theory (HQET),\cite{go} which is based on the
observation that in the limit $m_Q \rightarrow \infty$ the heavy quark decouples
from the light degrees of freedom. The heavy quark symmetry provides a good approximation 
for b hadrons since $m_b >> \Lambda_{QCD}$ and corrections obtained from a ${1 \over m_b}$
expansion are small. In the heavy quark limit, also
the spin of the heavy quark, $s_Q$, 
decouples from the orbital angular momentum of the system, $l$, and the spin of the
light quark(s) $s_q$. Both $s_Q$ and $ j= l \oplus s_q$ are separately conserved. Thus, states
are grouped into doublets bearing similar properties. The $B$
and $B^\star$ belong to the same doublet. 
The $l=1$ orbital excitations, frequently called $B^{\star\star}$'s, fall into two 
doublets. The $j={1 \over 2}$ states which include the scalar $B^\star_0$
and the axial vector $B^\star_1$ are broad decaying dominantly via S-wave to 
$B \pi$ and $ B^\star \pi$, respectively. 
The $j={3\over 2}$ states which include the axial vector $B_1$ and the tensor $B^\star_2$ 
are narrow. Their dominant decays proceed via D-wave to $B^\star \pi$ and $B^{(\star)} \pi$,
respectively. At the $Z^0$ the production of $l=0$ B-mesons 
versus $l=1$ B-mesons is favored by 7:3.  
According to spin counting $75\%$ of the $l=0$ B-states
are $B^\star$'s. The $B^{\star \star}$'s
decay strongly to $B^\star$'s or $B$'s with a ratio ranging between $1:1$ and $3:1$. 
Thus, B-mesons are the best laboratory to test HQET predictions. 

The crucial experimental tool for b-hadron spectroscopy is inclusive b hadron reconstruction.
For $b \bar b$ events selected via impact parameter tagging or via high $p_t$ muons, 
energy and momentum of the b-hadron are reconstructed, using either a rapidity algorithm 
(ALEPH,\cite{al1} DELPHI \cite{del1}) or  secondary vertex reconstruction (OPAL \cite{op1}). For 
events consistent with a b-hadron the Q-value defined by $Q_{BX} = m_{BX} - m_B - m_X$
is determined. For the majority of events in the final sample DELPHI \cite{del1} 
 e.g. achieves an energy resolution of $\sigma_E /E = 7 \%$ and angular resolutions of 
$ \sigma_\phi \simeq \sigma_\theta \simeq \rm 15\ mr$.

\section{Status of Pseudoscalar and Vector B Mesons}

The $B^+$ and $B^0$ masses have been measured rather precisely by CLEO,\cite{cl1} 
ARGUS \cite{arg1} and CDF.\cite{cdf1} 
The fits performed by the PDG \cite{pdg} yield: $ m_{B^+} = 5278.9 \pm 1.8$~MeV and 
$ m_{B^0_d} = 5279.2 \pm 1.8$~MeV.
The errors are dominated by a systematic uncertainty in determining the $e^+ e^-$ energy 
scale.\cite{cl1,arg1} 
The $B^0$ is heavier than the $B^+$ as expected. However, the observed
mass difference of $ m_{B^0_d} -  m_{B^+} = 0.35 \pm 0.29$ is consistent with zero. 
The $ B^0_s$ mass has been measured rather precisely by CDF \cite{cdf1} in 
the $B^0_s \rightarrow J/\psi \phi$ channel (see Figure~1).
The present $B^0_s$ mass measurements are depicted in Figure~2.
Including the LEP results \cite{al2,del2,op2}
the PDG mass fit yields $m_{B^0_s} = 5369.3 \pm 2.0$~MeV.
This constrains the $B^0_s - B^0_d$ mass splitting to $ 90 \pm 2.7$~MeV which 
is consistent with a quenched lattice QCD prediction of $107 \pm 13$~MeV.\cite{ma} 
The $B_c$ meson is not yet
observed. Searches have been conducted by all four LEP experiments 
\cite{al3,del3,l31,op3} and by CDF.\cite{cdf2} 
The channels studied include
$J/\psi \pi^+$, $J/\psi a_1^+$ and $J/\psi l \nu_l$.
Individual candidates are seen but
are consistent with the expected background. 
The most serious candidate is reported by 
ALEPH \cite{al3} in the $J/\psi \mu \nu_\mu$ channel
and has a mass of $m=5.96 \pm 0.25 \pm 0.19$~GeV.

\begin{figure}[h]
\vskip -3.0cm
\begin{center}
\setlength{\unitlength}{1cm}
\begin{picture}(12,5.0)
\put(0.,0.0)
{\mbox{\epsfysize=5.5cm\epsffile{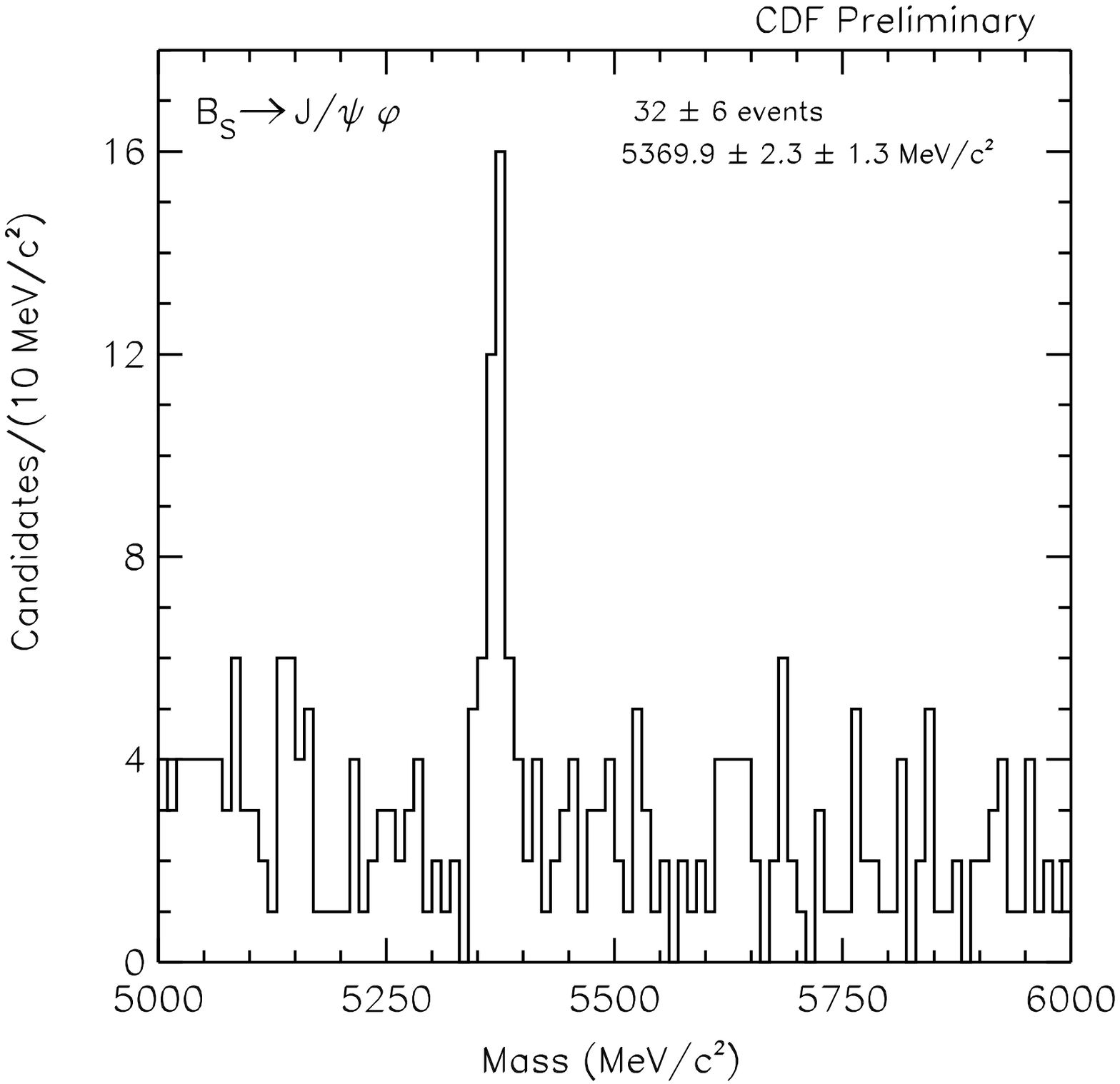}}\hspace*{-5mm}
\mbox{\epsfysize=7.3cm\epsffile{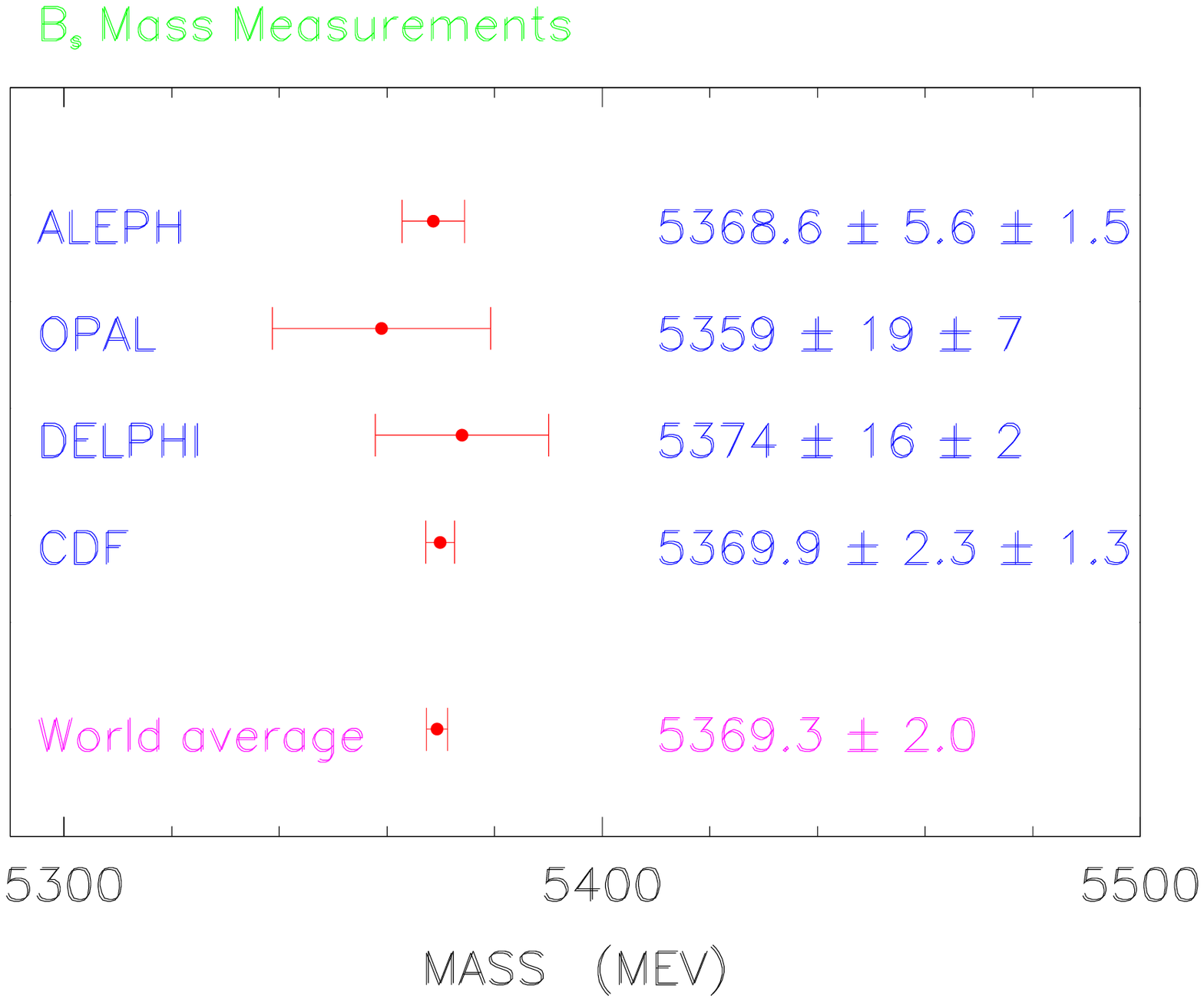}}}  
\end{picture}
\capttwo{The $J/\psi \phi$ invariant mass \hfill \break spectrum observed by CDF.
\label{fig:cdf_bs}}{Summary of all $B^0_s$ mass \hfill \break measurements.
\label{fig:bs_mass}}
\end{center}
\end{figure}

In the heavy quark expansion
hyperfine splittings (HFS) are proportional to ${1 \over m_Q}$. Thus, the $B$ and 
$B^\star$ are much closer spaced than the $D$ and $D^\star$. A quenched lattice calculation
e.g. yields $ \Delta m^{HFS}_B:= m_{B_{u,d}^\star} - m_{B_{u,d}} = 34 \pm 6$~MeV and  
$\Delta m^{HFS}_B:= m_{B_s^\star} - m_{B_s} = 27 \pm 17$~MeV.\cite{ma}
The small HFS permits
only electromagnetic (EM) transitions, of which $B^\star \ra \gamma B$ 
is the dominant one. For $B^\star$ reconstruction the low-energy photon needs to be detected. 
At LEP, however,  the photon energy is boosted, maximally up to 800~MeV.
L3 detects the photon directly in a crystal calorimeter, whereas the other 
LEP experiments reconstruct $e^+ e^-$ conversions. 
Typical energy and angular resolutions are $ \sigma_E/E = 1-2\%$ and 
$\sigma_{\theta,\phi}= 1-2\ mr$, respectively.

\begin{figure}[h]
\vskip -2.5cm
\begin{center}
\setlength{\unitlength}{1cm}
\begin{picture}(12,6.7)
\put(0.,0.0)
{\mbox{\epsfysize=6.5cm\epsffile{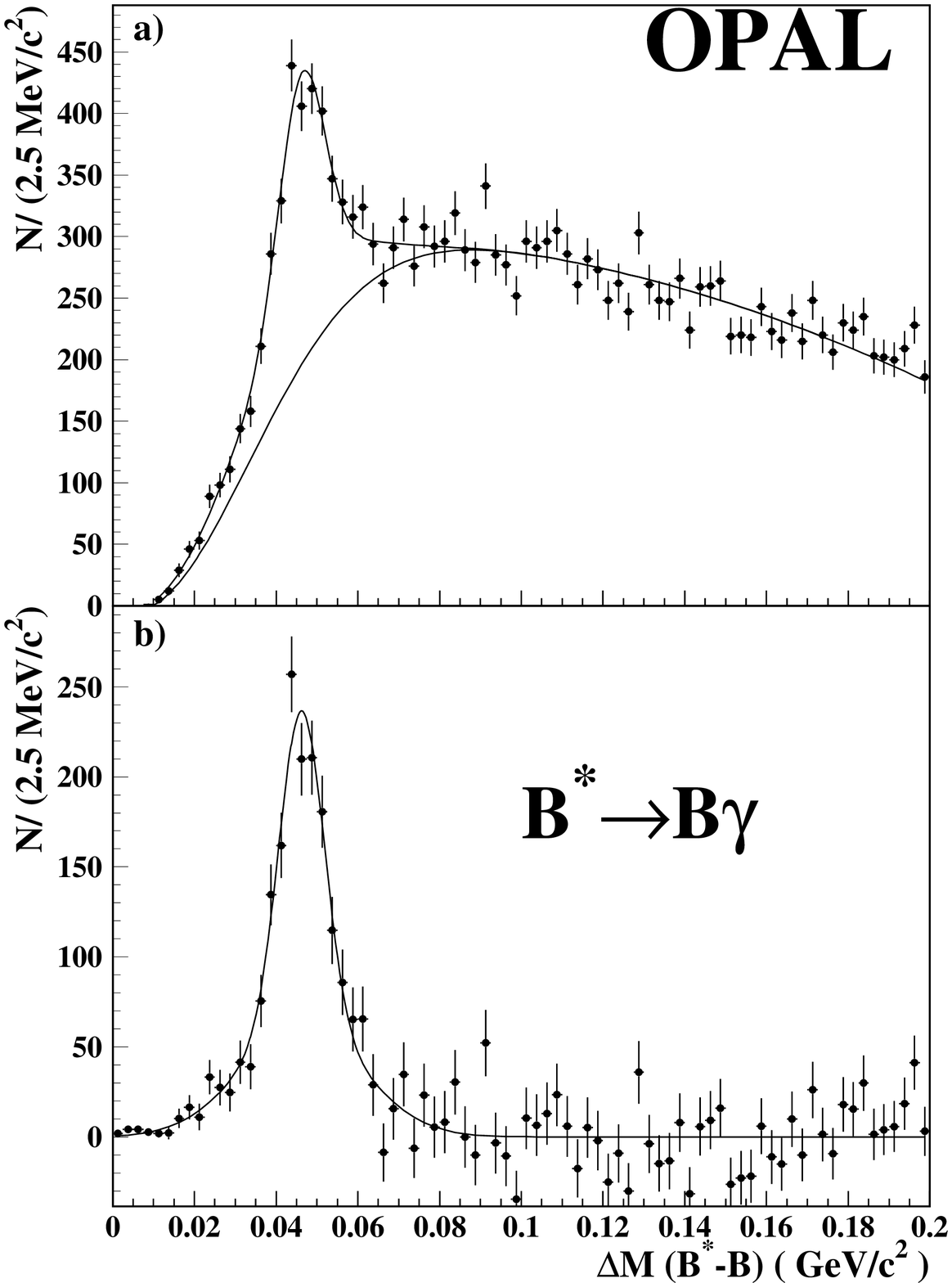}}\hspace*{-5mm}  
\mbox{\epsfysize=6.0cm\epsffile{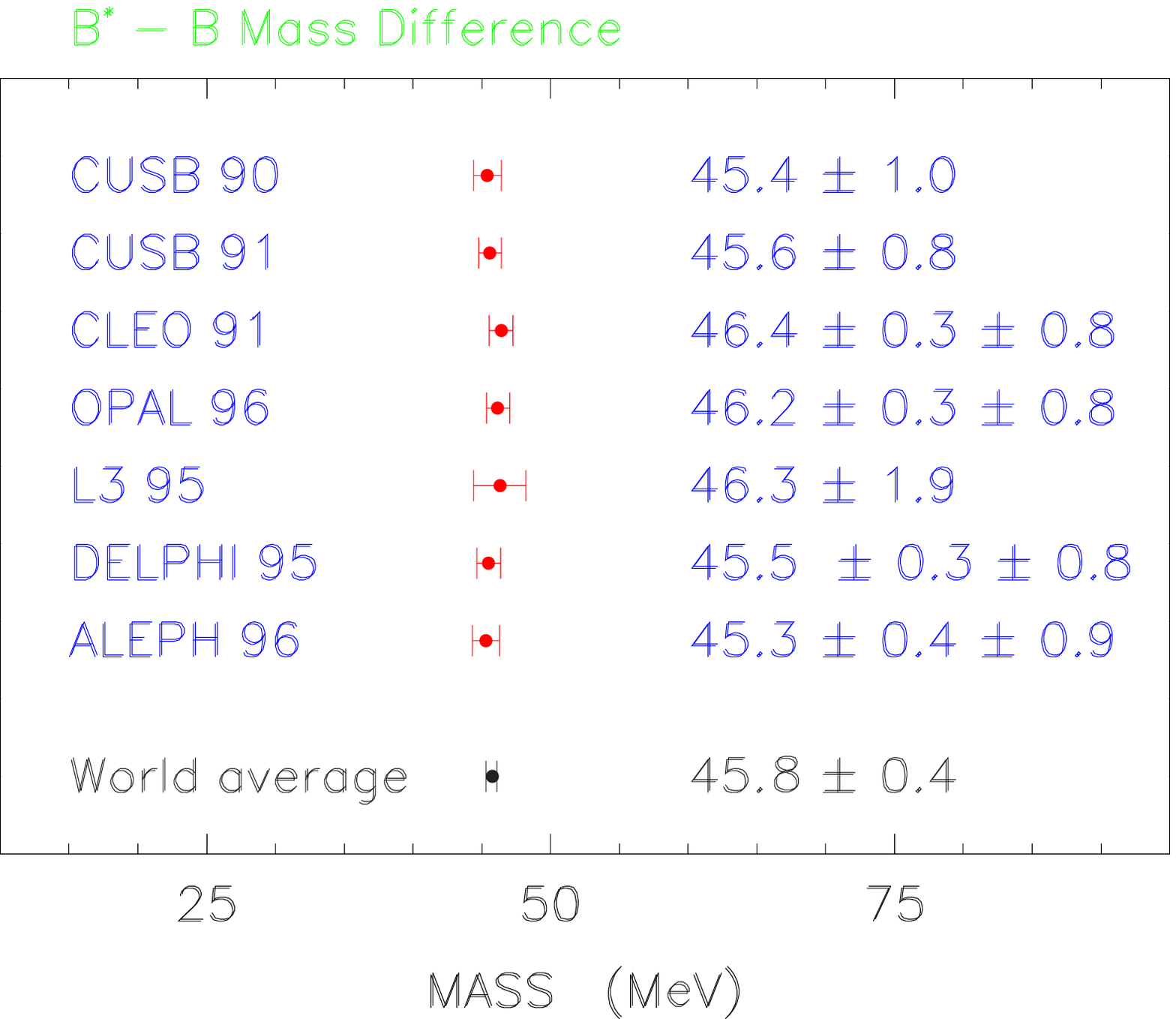}}}
\end{picture}
\capttwo{The $B^\star -B $ HFS \hfill \break observed by OPAL.
\label{fig:opal_bstar_1}}{Summary of all $B^\star -B $ \hfill \break 
HFS measurements.
\label{fig:bstar_mass}}
\end{center}
\end{figure}

\vskip 0.3cm
Figure~3 shows the $Q_{B\gamma}$-value distribution
measured by OPAL.\cite{op4}  Since the B-meson flavor is not identified, 
$ \Delta m^{HFS}_B$ includes contributions from $B^+$, $B_d^0$ and $B^0_s$. 
Figure~4 summarizes all $ \Delta m^{HFS}_B$
measurements from LEP \cite{op4,al4,del4,l32}
plus some old results from CLEO~II \cite{cl2} and CUSB.\cite{cu1}
The world average of the $B^\star - B$ hyperfine splitting 
is  $ \Delta m^{HFS}_B= 45.79 \pm 0.35$~MeV. 
DELPHI \cite{del4} also has set $95\%$ CL limits on 
$B^+ - B_d^0$ and $B_{u,d}^+ - B^0_s$
HFS differences, yielding: \break 
$\vert \Delta m^{HFS}_{B_u} - \Delta m^{HFS}_{B_d}
\vert < 6$~MeV and $\vert  \Delta m^{HFS}_{B_s} - \Delta m^{HFS}_{B_{u,d}}  
 \vert < 6$~MeV.

All LEP experiments have measured the relative $B^\star$ production 
cross section in $Z^0$ decays.\cite{op4,al4,del4,l32} 
Ignoring feed-down from $B^{\star \star}$'s the LEP average is 
${ \sigma_{B^\star} \over  \sigma_B +  \sigma_{B^\star} }= 0.748 \pm 0.004$.
This agrees with expectations from naive spin counting, since the adjustment
due to feed-down from $B^{\star \star}$'s is only a few \% effect
depending on assumptions made for 
$B^{\star \star}$ decays. ALEPH,\cite{al4} DELPHI~\cite{del4} and OPAL \cite{op4} also have 
measured the $B^{\star}$ polarization.
The observed helicity angle distribution is 
uniform, indicating that all helicity states are equally populated. 
The combined LEP result for the longitudinal helicity component is 
$ {\sigma_L \over \sigma_T + \sigma_L} =  0.33 \pm 0.04$.

Though the M1 transition is nearly $100\%$, higher order EM transitions like
the $B^\star$ Dalitz decays should also occur. The branching fraction independent of 
a form factor model is expected to be $ B(B^\star \ra B e^+ e^-) \simeq 0.466\%$.\cite{lan}
Since the $e^+$ and $e^-$ momenta are below 100~MeV, electron identification 
and tracking in the vertex detector are essential. DELPHI has combined Dalitz 
pairs originating from the primary vertex with a B candidate.\cite{delb} The resulting 
$Q_{Be^+e^-}$ distribution plotted in Figure~5 shows a peak at the expected 
$\Delta m_B^{HFS}$ value. 
The $B^\star$ Dalitz decay rate normalized to that of the M1 transition 
is measured to be:
$ \Gamma(B^\star \ra B e^+ e^-)/ \Gamma(B^\star \ra B \gamma) 
= ( 4.7 \pm 1.1 \pm 0.9) \times 10^{-3}$.

\section{Status of Orbitally-Excited B-Mesons}

HQET predicts two narrow and two broad $B^{\star \star}$  states 
similarly as in the $D^{\star \star}$ system. Using the heavy quark expansion 
Eichten, Hill and Quigg determine the masses and total widths of the two $j={3\over 2}$ 
$B^{\star \star}_{u,d}$ states to be $m_{B_1} = 5759$~MeV,
$m_{B^\star_2} = 5771$~MeV, $\Gamma_{B_1} = 21$~MeV and 
$\Gamma_{B^\star_2}=25$~MeV, respectively.\cite{eic1}  
A recent calculation by Falk and Mehen based on the heavy flavor expansion 
yields masses of $m_{B_1} = 5780$~MeV and 
$m_{B^\star_2} = 5794$~MeV.\cite{fa1} 
The masses of the $j={1 \over 2}$ states are expected to lie about 100~MeV lower than
those of the $j={3\over 2}$ states. 

ALEPH,\cite{al5} DELPHI \cite{del5,del6} and OPAL \cite{op5} have analyzed 
single $\pi^+$ transitions using inclusive 
B reconstruction methods. The Q-value distribution measured by DELPHI is plotted in 
Figure~6. A broad structure is observed at $m = 5734 \pm 5 \pm 17$~MeV.
A decomposition into individual $j={3\over 2}$ and  $j={1\over 2}$ states is presently 
not conclusive. Similar observations have been found by the other LEP experiments. 
A summary of all mass measurements is shown in Figure~7. Note that the masses 
from OPAL and ALEPH shown here have been shifted up by 31~MeV~\footnote{We 
have assumed that $B^\star \pi$ versus $B \pi$ transitions
are enhanced by $2\pm 1 : 1$, by considering various scenarios for 
$B^{\star \star}$ production and decay.\cite{fein}} 
to account for dominant contributions from $B^\star \pi$ 
transitions.\cite{del6}
The combined LEP result for the $B^{\star \star}_{u,d}$ mass
is $m(B^{\star \star}_{u,d}) = 5722 \pm 8$~MeV. 
This is lower than the mass predictions for $j={3\over 2}$ states, thus leaving room for
contributions from the $j={1 \over 2}$ states. 
ALEPH, in addition, has performed an exclusive analysis in the $B \pi$ channel.\cite{al6}
A significant narrow structure is seen at $m=5703 \pm 14$~MeV. 
The resolution of $\sigma = 28 \pm ^{18}_{14}$ MeV would permit
contributions from both $j={3 \over2}$ states. However,
even after the +31~MeV shift,\cite{fein}
the mass is still too low to agree with a DELPHI measurement 
obtained in the $B^{(\star)} \pi \pi$ final state (see below).

\begin{figure}[thb]
\vskip -1.0cm
\phantom{10mm}
\begin{center}
\setlength{\unitlength}{1cm}
\begin{picture}(12,7.0)
\put(0.,0.0)
{\mbox{\epsfysize=6.5cm\epsffile{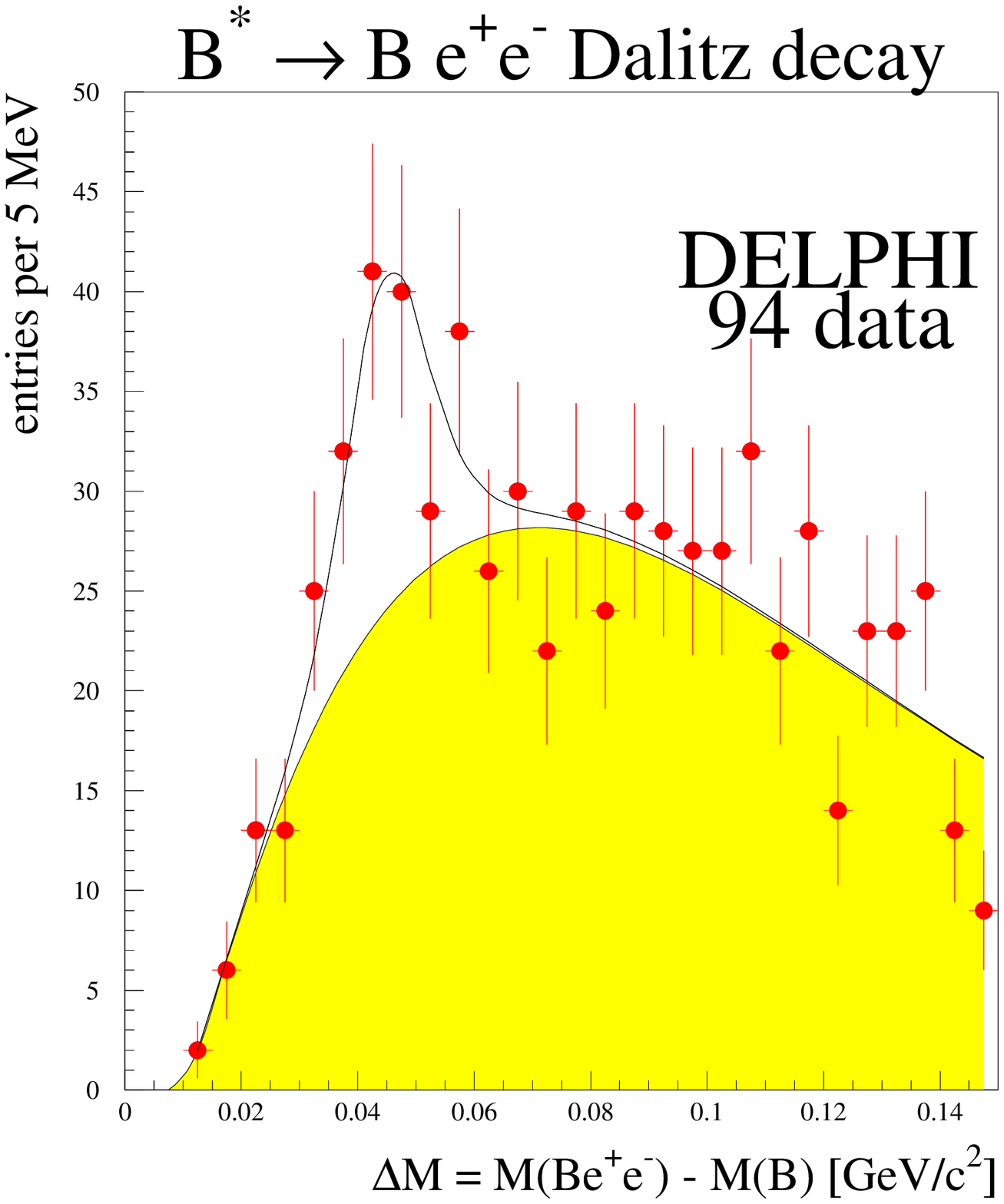}}\hspace*{-5mm}  
\mbox{\epsfysize=7.5cm\epsffile{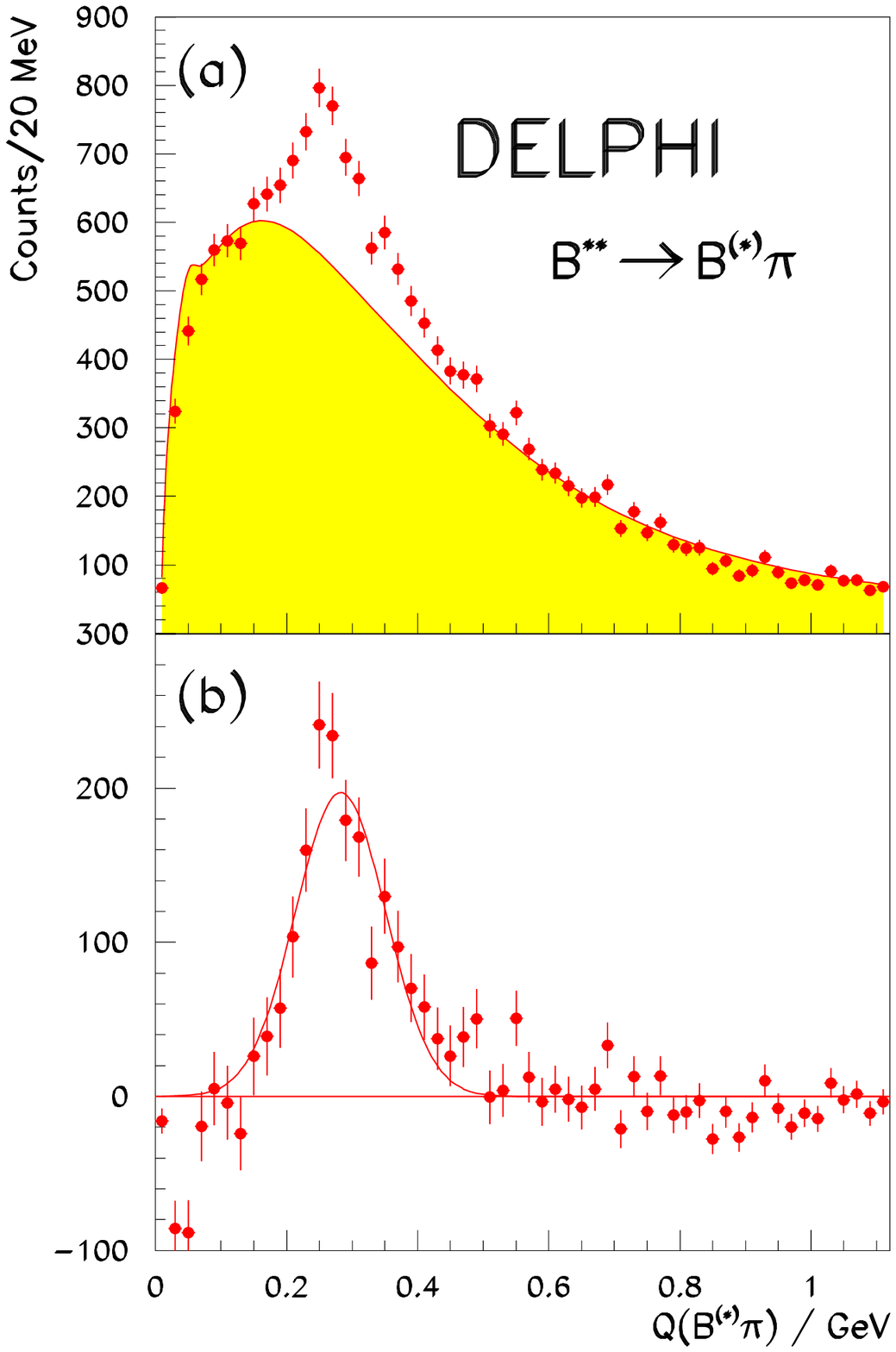}}}
\end{picture}
\capttwo{The Q-value distribution \hfill \break for  $B^\star$ Dalitz decays.
\label{fig:bdalitz}}{The Q-value distribution \hfill \break for $B^{(\star)} \pi$
final states.
\label{fig:q_bdstar}}
\end{center}
\end{figure}

\vskip 0.5cm
 
The decay angle distribution of the $\pi$ in the $B^{\star \star}$ rest frame provides information
on the helicity distribution of the light quark system.\cite{fa2}
DELPHI has observed a uniform decay
angle distribution,\cite{del5}
which implies that the maximally allowed helicity components of the light
quark system are not suppressed. This is surprising since ARGUS has observed the opposite
for $D^\star_2$ decays.\cite{arg2} Assuming that the contribution of decays from the
$j={1 \over 2}$ states,
which produce a non-uniform decay angle distribution, is small, a fraction of
$w_{{3\over 2}}= 0.53 \pm 0.07 \pm 0.10$ is measured for the helicity $j=\pm{3 \over 2}$ 
components.\footnote{The S-wave decays $B^\star_0 \ra B \pi$ and
$B^\star_1 \ra B^\star \pi$ are expected to be dominant.}

The masses predicted by Eichten, Hill and Quigg for the narrow $B^{\star \star}_s$ states  
are $m_{B_{s1}} = 5849$~MeV and $m_{B^\star_{s2}} = 5861$~MeV.\cite{eic1} Predictions by 
Falk and Mehen are again higher, yielding
$m_{B_s1} = 5886$~MeV and $m_{B^\star_{s2}} = 5899$~MeV.\cite{fa1}
Since the $B^{\star \star}_s - B_{u,d}$ mass difference is larger than the kaon mass, the
dominant transitions are $B^{\star \star}_s \ra K B_{u,d}^\star$.

Using the inclusive analysis techniques DELPHI has studied $B^{(\star )} K^\pm$ final 
states.\cite{del6}
DELPHI is rather suited for analyzing such channels because of their excellent kaon identification
over a wide momentum range.\cite{del7} The resulting Q-value distribution 
depicted in Figure~8 shows two 
narrow structures at $70 \pm 4 \pm 8$~MeV and $142 \pm 4 \pm 8$~MeV. 
Their widths are slightly smaller
than the observed resolution. Assuming that the upper peak stems from the transition
$B_{s1} \ra B^\star K$ and the lower peak stems from   $B^\star_{s2} \ra B K$, 
masses of  $ m_{B_{s1}} = 5888 \pm 4 \pm 8$~MeV and 
$ m_{B^\star_{s2}} = 5914 \pm 4 \pm 8$~MeV have been obtained.
The mass splitting is $ m_{B^\star_{s2}}- m_{B_{s1}} = 26 \pm 6 \pm 8$~MeV. Both the masses
and the splitting are higher than the HQET predictions. 
In addition, upper limits have been set on the widths, 
yielding $ \Gamma_{B_{s1}} < 60$~MeV and 
$ \Gamma_{B^\star_{s2}} < 50$~MeV at $95\%$ CL, respectively. The production 
cross sections for $B_{s1}$ and $B^\star_{s2}$ states with respect to that of
$B^{\star \star}_{u,d}$ has been measured to be: 
$ (\sigma_{B_{s1}} + \sigma_{B^\star_{s2}}) /\sigma_{B^{\star\star}_{u,d}}
= 0.142 \pm 0.028 \pm 0.047$. OPAL \cite{op5} has also studied this channel, observing a
$\Gamma = 47$~MeV broad structure at $m= 5853\pm 15$~MeV, which again needs to be shifted upward 
by $\sim 31$~MeV to account for dominant $B^\star K$ transitions.

\begin{figure}[thb]
\vskip -2.5cm
\phantom{10mm}
\begin{center}
\setlength{\unitlength}{1cm}
\begin{picture}(12,8.0)
\put(0.,0.0)
{\mbox{\epsfysize=5.0cm\epsffile{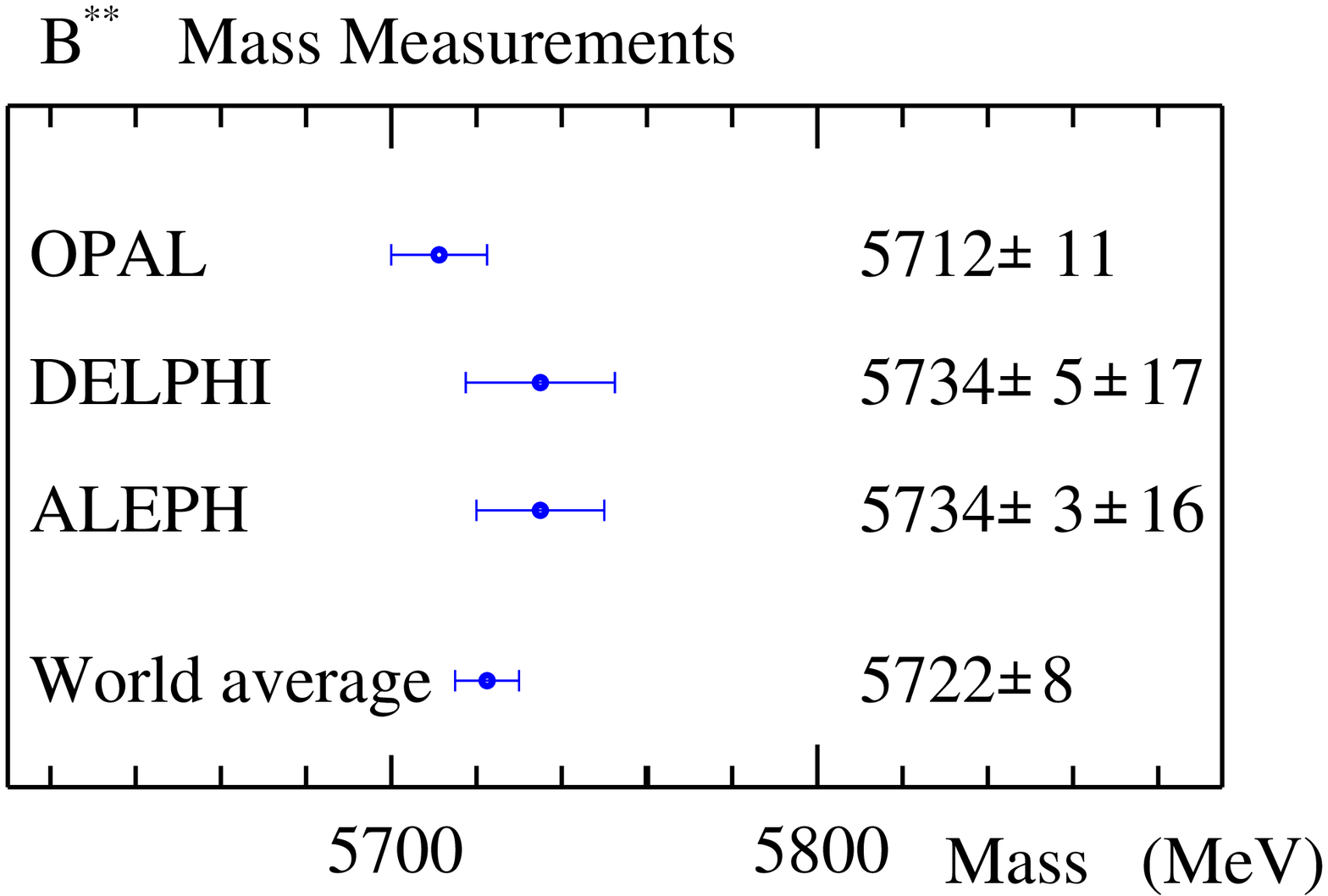}}\hspace*{-5mm}  
\mbox{\epsfysize=7.0cm\epsffile{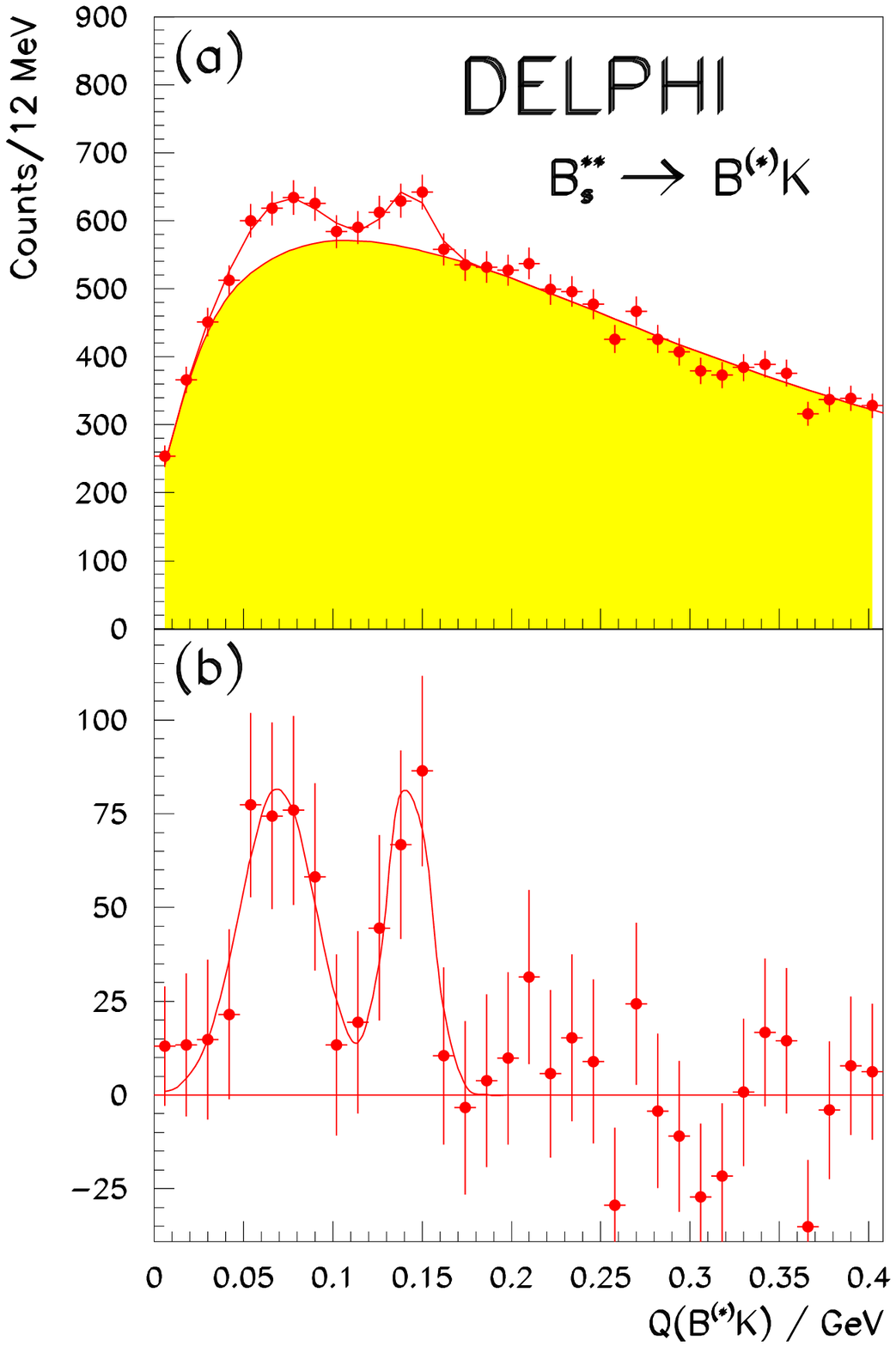}}}
\end{picture}
\capttwo{Summary of all inclusive $B^{\star \star}$ \hfill \break mass
 measurements.
\label{fig:m_bdstar}}{The Q-value distribution \hfill \break for $B K^\pm$ final states.
\label{fig:bsdstar}}
\end{center}
\end{figure}

\section{First Observation of Radially Excited B Mesons}

Radial excitations of D mesons and B mesons should exist similarly as those of
$ c \bar c$ and $b \bar b$ states.\cite{go}
A QCD inspired relativistic quark model predicts the masses of the
2S pseudoscalar and vector states to lie at $5900$~MeV and $5930$~MeV, 
respectively.\cite{ko}
DELPHI has extended the inclusive analysis to $\pi^+ \pi^-$ transitions 
to look for such states.\cite{eig1,del8} 
For $b \bar b$ events with a $\pi^+ \pi^-$ pair from the primary vertex where both pions have  
large rapidities ($\eta >2.5$) and are in the same hemisphere as the B candidate, 
the variable, 
$Q_{B\pi\pi}=m_{B^{(\star)} \pi^+ \pi^-} - m_{B^{(\star)}} - 2 m_{\pi^{\pm}}$, was 
determined. This selection is $52\pm3\%$ efficient and has a purity of $80 \pm 4\%$.
The resulting Q-value distribution displayed in Figure~9 shows two narrow
structures, one at $Q_{B^{(\star)}\pi\pi}=301\pm4\pm10$~MeV containing $56\pm13$ events 
and a second at $Q_{B^{(\star)}\pi\pi}=220 \pm 4 \pm10$~MeV containing $60 \pm 12$ events. 
The corresponding measured
resolutions, $\sigma=12\pm3$~MeV and $\sigma=15\pm3$~MeV, are compatible with the detector 
resolution, implying that their natural widths must be narrow. Thus, the two broad 
$j = {1 \over 2}$ orbital excitations cannot contribute significantly here. 

Figure~10 shows all allowed transitions for the 2S states and 1P states.
The non-suppressed $\pi$ and $\pi\pi$ transitions of the narrow $j=3/2$ 
P states are:
$B_1~\ra~B^\star~\pi$~(D-wave), $ B_1~\ra~B~\pi\pi$ \&
$B_1\rightarrow B^\star \pi\pi$ (P-wave); 
$B_2^\star~\rightarrow~B\pi$~\&~$B_2^\star\rightarrow~B^\star\pi$  (D-wave), and 
$B_2^\star\rightarrow B^\star \pi\pi$ (P-wave). 
The corresponding transitions of the 2S states are: 
$B^\prime \ra B^\star \pi$  (P-wave), $B^\prime \ra B_0^\star \pi$ \&
$B^\prime \ra B \pi\pi$  (S-wave);  
$B^{\star\prime} \ra B \pi$ \&
$B^{\star\prime}\ra B^\star \pi$ (P-wave), and $B^{\star\prime} \ra B_1 \pi$ 
\& $B^{\star\prime} \ra B^\star \pi\pi$ (S-wave).
$\rho$ transitions
are suppressed by phase space. Since the mass resolution is smaller than 
$ \Delta m^{HFS}_B$, we can exclude that a single excited state decays to $B \pi \pi$ 
and $B^\star \pi\pi$ simultaneously. In that case two peaks separated by $ \Delta m^{HFS}_B$
should have been visible. It is, however, possible that the two peaks originate from two 
closely spaced  excited states, where the heavier decays to $ B^\star \pi \pi$
and the lighter to $B \pi \pi$.

\begin{figure}[thb]
\vskip -2.5cm
\phantom{10mm}
\begin{center}
\setlength{\unitlength}{1cm}
\begin{picture}(12,7.5)
\put(0.,0.0)
{\mbox{\epsfysize=6.0cm\epsffile{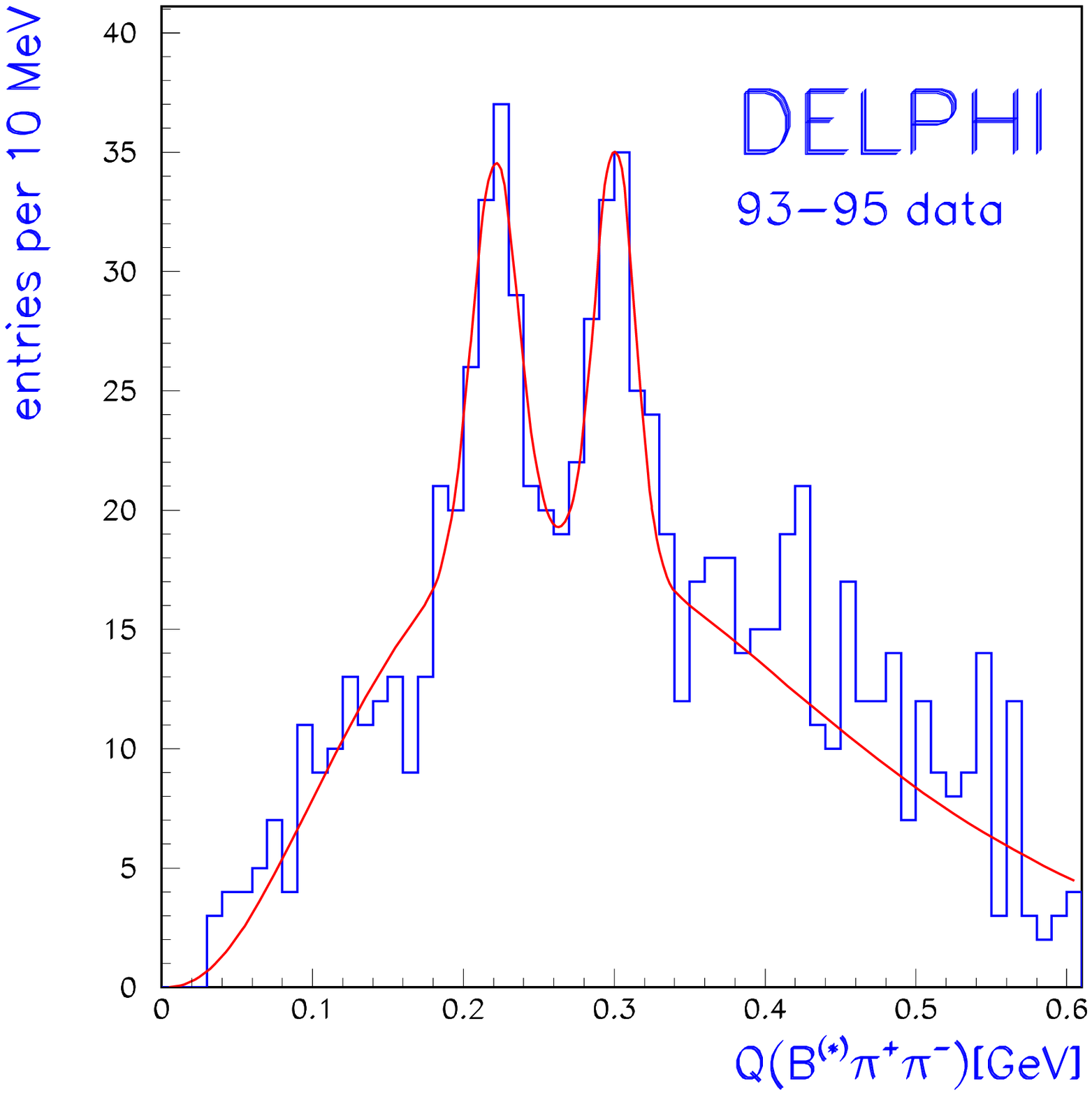}}\hspace*{-5mm}  
\mbox{\epsfysize=5.5cm\epsffile{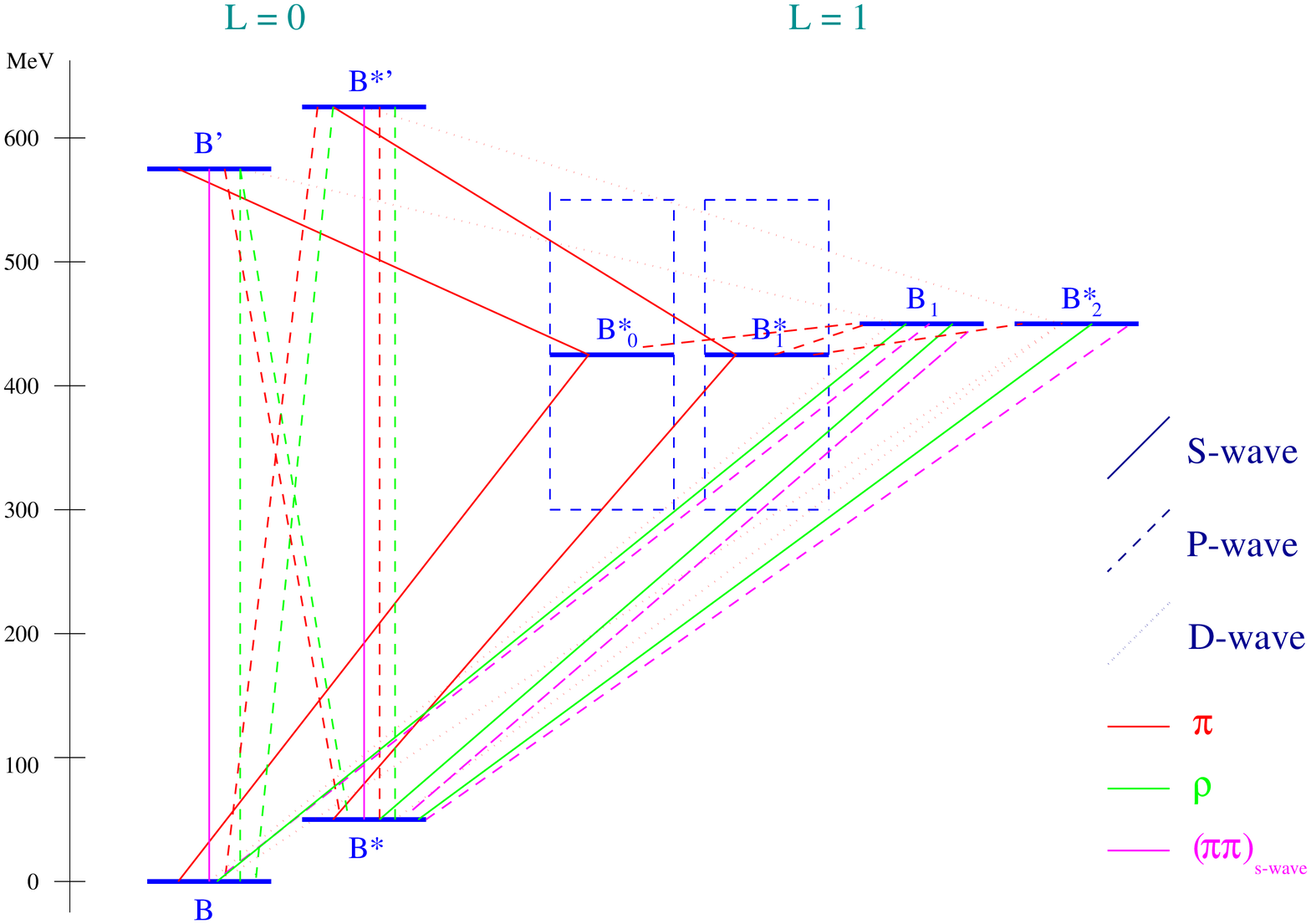}}}
\end{picture}
\capttwo{The Q-value distribution \hfill \break for $B^{(\star)}
 \pi^\pm \pi^\mp$ final states.
\label{fig:radial}}{B-meson level diagram \hfill \break with expected hadronic transitions.
\label{fig:radial_ld}}
\end{center}
\end{figure}

The lower peak most likely stems from the P-wave transitions,
$B_1 \rightarrow B \pi^+ \pi^-$ with a possible contribution from 
$B_2^\star\rightarrow \pi^+\pi^- B^\star$. 
Denoting the mass splitting of the $j={ 3 \over 2} $ states by
$ \Delta m_{B_{3/2}}:=m_{B^\star_2} - m_{B_1}$ and the fraction of 
$B_2^\star $ decays by f, we can parametrize the masses of the narrow states 
by:
$ m_{B_1} = 5778 + f \cdot (\Delta m_B^{HFS} -  \Delta m_{B_{3/2}}) \pm 
11$~MeV and $m_{B^\star_2} = 5824 - (1-f) \cdot (\Delta m_B^{HFS} -  
\Delta m_{B_{3/2}}) \pm 11$~MeV. 
These mass estimates are 
consistent with the heavy quark predictions for $j={3\over 2}$ states,
but they are higher than the mass of the broad structure observed in Figure~6. 
This implies that 
$j={1\over 2}$ states contribute significantly there. 
Assuming that $0 \leq \Delta m_{B_{3/2}} \leq \Delta m_B^{HFS}$ as in the 
D-system, we can set bounds of $m_{B_1} > 5756$~MeV and $m_{B^\star_2} < 
5846$~MeV $@ 95\%$ CL, which are in conflict with the exclusive 
$B^{(\star)} \pi$ ALEPH result.\cite{al6}

The upper peak has to originate from states which lie $\geq 80$~MeV above 
the $B_1$. 
The most likely interpretation is that this peak stems from $\pi \pi$ transitions of the 2S 
radial excitations: $B^{\prime} \rightarrow B \pi \pi$ and 
$B^{\star\prime} \rightarrow B^\star \pi \pi$. The S-wave $\pi \pi$ transitions are expected 
to be dominant, though two successive $\pi$ transition via the broad
$ j={1 \over2}$
orbital excitations are kinematically allowed. However, more detailed studies are needed
to clarify this issue.
Though single $\pi$ transitions $B^{(\star)\prime} \rightarrow B^{(\star)} \pi$ are allowed, 
they should be suppressed because of nodes in the radial wave functions, which lead to
cancelations in the overlap integral.\cite{ko} Such cancelations have been observed in 
$\rho^\prime \ra \pi \pi$ \cite{kau} and $ \psi(4040) \ra D \bar D$ \cite{ya} decays. 
Nevertheless, the observed Q-value distribution  
for $B^{(\star)} \pi$ final states actually has room for such transitions. 
Assuming that the production of
$B^{\star\prime}$ to $B^\prime$ is similar to that of $B^\star$
to $B$, we obtain the following mass estimates for the 2S states:
$m_{B^\prime}=5859+{3\over 4}(\Delta m_B^{HFS}- \Delta m^{HFS}_{B^\prime})
\pm 12$~MeV and \break
$m_{B^{\star\prime}}=5905-{1\over 4}(\Delta m_B^{HFS}- 
\Delta m^{HFS}_{B^\prime})\pm 12$~MeV.
Here $ \Delta m^{HFS}_{B^\prime}$ denotes the HFS of the radially excited states. 
These values are consistent with predictions from a QCD inspired relativistic quark model,
thus supporting the interpretation of observing 2S radial excitations.\cite{ko}
A preliminary estimate of the production cross sections from the observed signal
yield is: 
$\sigma(b \rightarrow B^\prime + B^{\star\prime})/\sigma(b \ra all) = 0.5\% - 4\%$. 
The branching ratio for $B_1 \rightarrow B \pi \pi$ is of the order of $2\% - 10\%$.

\section{Status of b Baryons}

The $\Lambda_b$ is clearly established since a recent CDF measurement in the 
$\Lambda J/\psi$ channel.\cite{cdf3} The $\Lambda J/\psi$ invariant mass peaks at
$m_{\Lambda_b} = 5621 \pm 4 \pm 3$~MeV. Previously, ALEPH \cite{al7} and DELPHI \cite{del9} had
observed a few candidates in the $\Lambda_c^\pm \pi^\mp$ channel. All
mass measurements are summarized in Figure~11. The present world average for
the $\Lambda_b$ mass is $m_{\Lambda_b} = 5624 \pm 9$~MeV.\cite{pdg}

\begin{figure}[thb]
\vskip -2.5cm
\phantom{10mm}
\begin{center}
\setlength{\unitlength}{1cm}
\begin{picture}(12,8.5)
\put(0.,0.0)
{\mbox{\epsfysize=5.0cm\epsffile{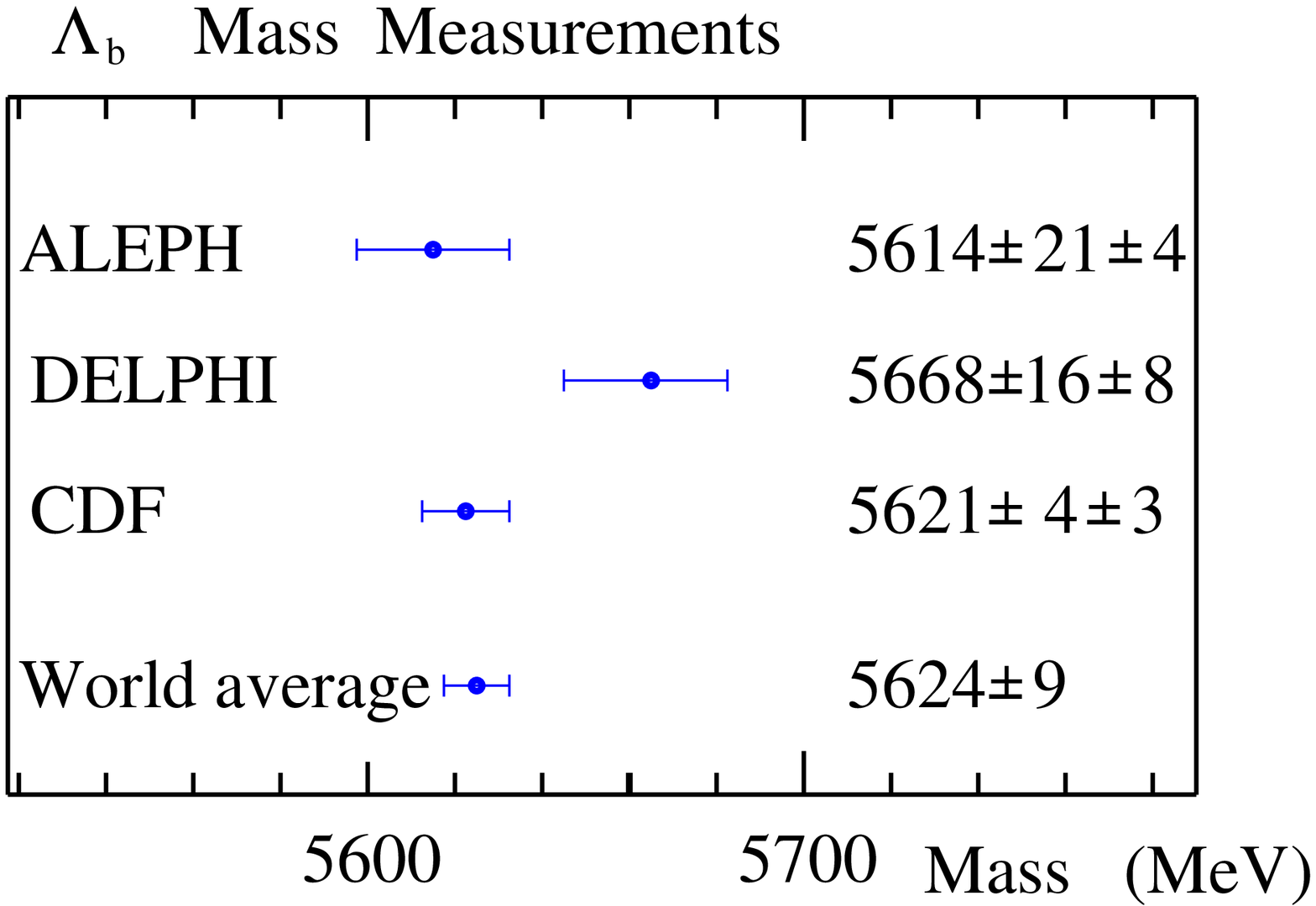}}\hspace*{-5mm}  
\mbox{\epsfysize=7.5cm\epsffile{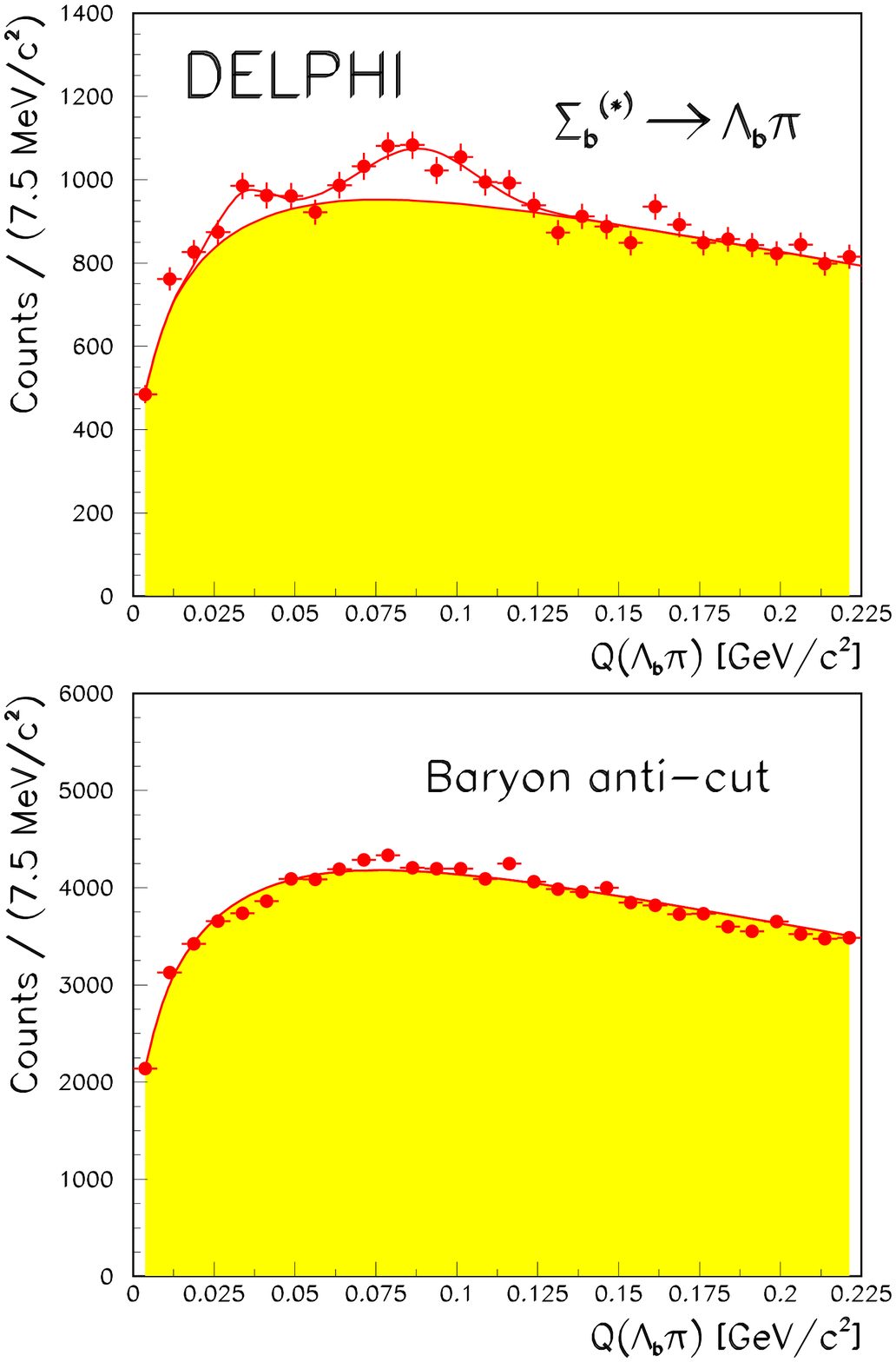}}}
\end{picture}
\capttwo{Summary of $\Lambda_b$ mass \hfill \break measurements.
\label{fig:radial}}{The Q-value distribution \hfill \break for $\Lambda_b \pi $ final states.
\label{fig:radial_ld}}
\end{center}
\end{figure}

\vskip 0.5cm

Using the heavy quark expansion in combination with the observed $\Sigma^\star_c - \Sigma_c$ HFS provides a prediction for the $\Sigma^\star_b~-~\Sigma_b$~HFS 
of $\Delta m^{HFS}_{\Sigma_b} = 22$~MeV.\cite{fa3} \break
DELPHI \cite{dela} has looked for b-flavored baryons using the 
inclusive analysis techniques.
Baryon enrichment is obtained by selecting fast p's, n's and $\Lambda$'s. For events 
consistent with a $\Lambda_b$, a pion from the primary vertex is added to determine the variable
$ Q_{\Lambda_b \pi} = m_{\Lambda_b \pi} -m_{\Lambda_b} - m_\pi$.
The resulting distribution depicted in Figure~12
reveals two structures at $ Q_{\Lambda_b \pi} = 33 \pm 3 \pm 8$~MeV and 
$Q_{\Lambda_b \pi}= 89 \pm 3 \pm 8$~MeV.    
Interpreting these as transitions from the $\Sigma_b$ and $\Sigma_b^\star$,
mass differences of $m_{\Sigma_b} - m_{\Lambda_b} = 173 \pm 3 \pm 8$~MeV
and $m_{\Sigma^\star_b} - m_{\Lambda_b} = 229 \pm 3 \pm 8$~MeV are determined. 
Within errors they are consistent with quark model predictions \cite{ron} 
yielding $m_{\Sigma_b} - m_{\Lambda_b} = 200 \pm 20$~MeV and 
$m_{\Sigma^\star_b} - m_{\Lambda_b} = 230 \pm 20$~MeV, respectively.   
The measured HFS of  $ \Delta m _{\Sigma_b}^{HFS} = 56 \pm 15$~MeV is in conflict
with HQET predictions.\cite{fa3} 
This measurement needs to be checked, since presently it
cannot be ruled out that either a transition from a different state is seen or that for one of the
structures the observed mass is shifted 
due to contributions from another transition. It is worthwhile to
note that the lower peak is narrower than the higher peak. This is supportive 
for a more complex interpretation. 
To clarify this issue
DELPHI plans to redo the analysis with reprocessed data, which
show significant improvements in track reconstructions and thus achieve improved efficiencies
and momentum resolutions.

Assuming that the two peaks stem from $\pi$ transitions of the $\Sigma_b$ and 
$\Sigma^\star_b$, DELPHI measures a relative production cross section 
of $(\sigma_{\Sigma_b} + \sigma_{\Sigma^\star_b})/ \sigma(b \ra all) = 4.8 \pm 0.6 \pm 1.5 \%$.
The fraction originating from $\Sigma$ baryons is  $24 \pm 6 \pm 10 \%$.

DELPHI has also measured the helicity angle distribution of the $\pi$ in the 
$\Sigma_b^\star$ rest frame. A fit to the Falk Peskin model \cite{fa2}
yields $ w_1 = 0.36 \pm 0.30 \pm 0.030$ for the helicity $h=\pm1$ component
of the light quark system, indicating that these states are suppressed. 
According to Falk and Peskin~\cite{fa2} large $\Sigma_b$ and
$\Sigma^\star_b$ rates in combination with a
suppression of helicity $\pm 1$ states lead to a substantial reduction
of the $\Lambda_b$ polarization in $Z^0$ decays.
This has in fact been observed by ALEPH,\cite{al8} measuring $ P(\Lambda_b) 
= -0.26^{+0.25}_{-0.20}$$ ^{+0.13}_{-0.12}$.

\begin{figure}[thb]
\vskip -2.5cm
\phantom{10mm}
\begin{center}
\setlength{\unitlength}{1cm}
\begin{picture}(12,7.9)
\put(0.,0.0)
{\mbox{\epsfysize=6.0cm\epsffile{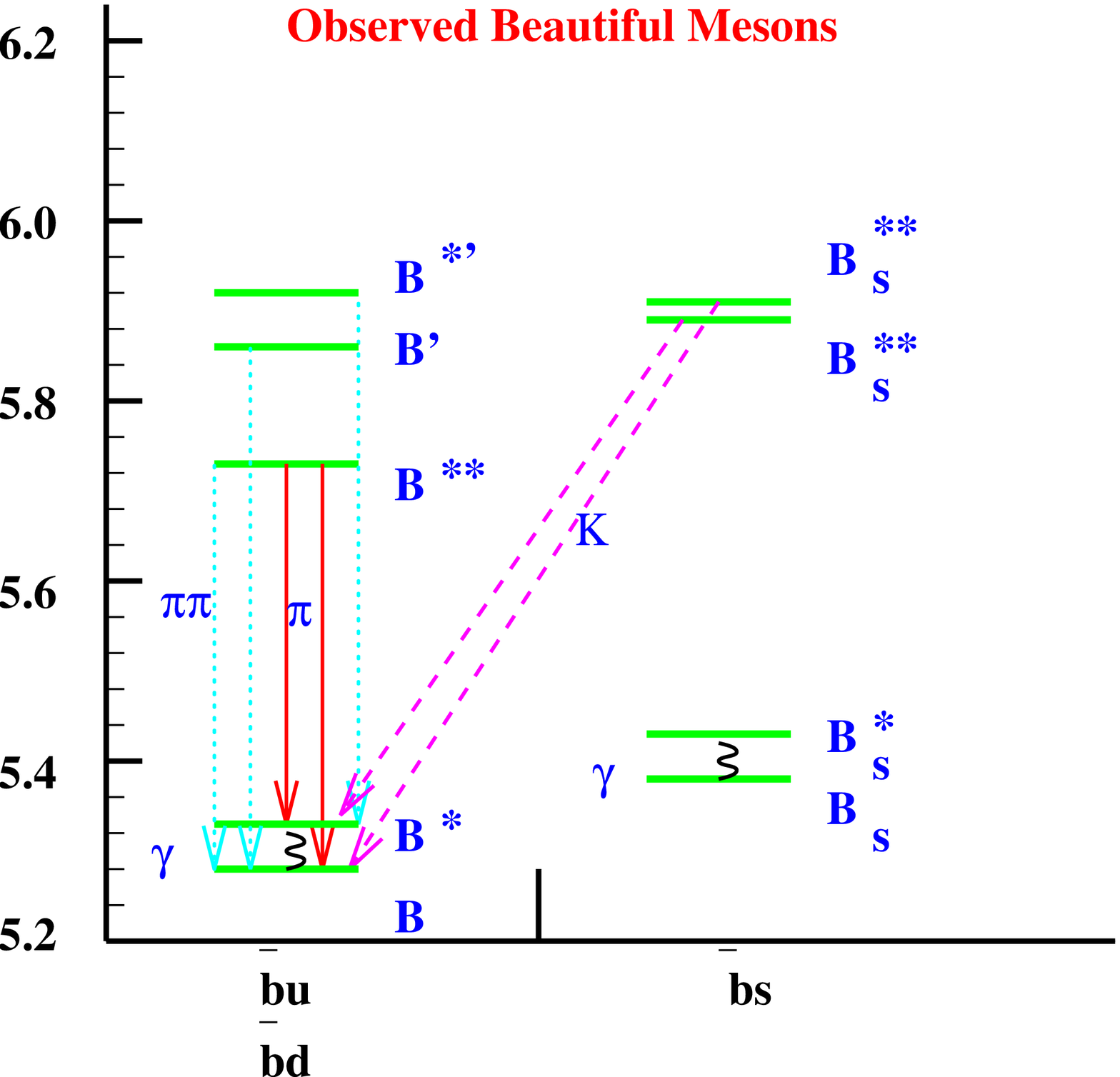}}\hspace*{-5mm}  
\mbox{\epsfysize=6.0cm\epsffile{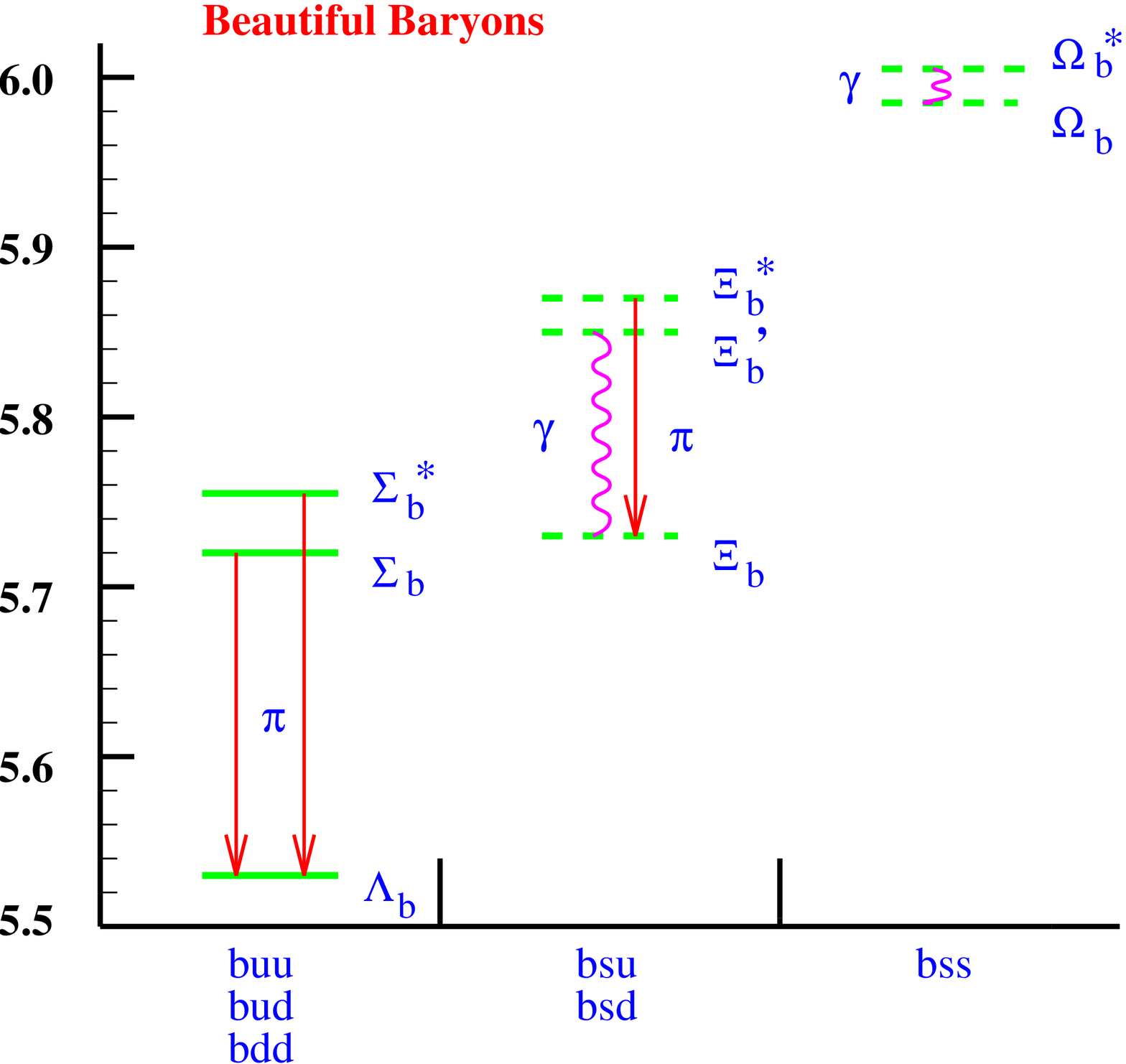}}}
\end{picture}
\capttwo{Observed B-mesons masses and \hfill \break transitions.
\label{fig:radial}}{b-baryon states and \hfill \break transitions. Observed states
\hfill \break are shown by solid lines. 
\label{fig:radial_ld}}
\end{center}
\end{figure}

\section{Summary}

The knowledge of the B meson sector has improved over the past few years.
The present status is summarized in Figure~13. 
Precisely measured masses exist for all pseudoscalar and vector B meson ground states 
except for those in the $B_c$ system, which has not been detected yet.
Evidence is found for orbital excitations, but only in the $B^0_s$ system it was possible 
to isolate two separate narrow states. Cross section measurements agree with expectations and
decay angle distributions indicate that helicity $j=\pm {3 \over 2}$ states are not suppressed.
While presently there is no evidence for orbital excitations with $L>1$,
first evidence is found for the 2S $B_{u,d}$ radial excitations.

States and transitions in the b-baryon sector are summarized in 
Figure~14. 
The knowledge is still rather poor here. Only the $\Lambda_b$ is 
well-established. The $\Sigma_b$ and $\Sigma_b^\star$ may be observed but the HFS is
in conflict with HQET predictions. 
Thus, these measurements need confirmation. 
The mass of the $\Xi_b$ is unknown, though its lifetime has been measured.\cite{pdg}
So far no other b-flavored baryon has been identified.

\section{Acknowledgments}
This work has been supported by the Research Council of Norway. 
I would like to acknowledge the DELPHI collaboration for support. Special thanks goes to 
M. Feindt, Ch. Weiser and Ch. Kreuter for fruitful discussions.

\end{document}